# An Adaptable Maturity Strategy for Information Security

[1,2] Gliner Dias Alencar, [2] Hermano Perrelli de Moura, [2] Ivaldir Honório de Farias Júnior,
[2] José Gilson de Almeida Teixeira Filho
[*1, First Author, Corresponding Author] *Brazilian Institute of Geography and Statistics – IBGE*
[2]*Federal University of Pernambuco in Brazil – UFPE,*
*gda2@cin.ufpe.br, hermano@cin.ufpe.br, ihfj@cin.ufpe.br, jgatf@cin.ufpe.br*

## Abstract

*The lack of security in information systems has caused numerous financial and moral losses to several organizations. The organizations have a series of information security measures recommended by literature and international standards. However, the implementation of policies, actions, and adjustment to such standards is not simple and must be addressed by specific needs identified by the Information Security Governance in each organization. There are many challenges in effectively establishing, maintaining, and measuring information security in a way that adds value.* Those challenges demonstrate a need for further investigations which address the problem. *This paper presents a strategy to measure the maturity in information security aiming, also, to assist in the application and prioritization of information security actions in the corporate environment. For this, a survey was used as the main methodological instrument, reaching 157 distinct companies. As a result, it was possible to classify the ISO/IEC 27001 and 27002 controls in four stages according to the importance given by the companies. The COBIT maturity levels and a risk analysis matrix were also used. Finally, the adaptable strategy was successfully tested in a company.*

**Keywords**: *Information Security, Maturity, Prioritization, Governance, Management*

## 1. Introduction

Most of the research in the field of information security has focused, as a priority, on the development, improvement, and application of technical aspects in the systems, networks and physical security, for example, [1] [2] [11]. However, the single application of technology is not sufficient for the treatment of information security. For the new challenges, it is also necessary to address other areas, looking at information security more broadly [3]. Within this holistic view are highlighted the studies that address the aspects and human influences in information security [4][5] and the area related to processes, procedures, and controls. They cover management, governance, auditing, compliance, information security policy, and maturity, the latter being the focus of the present study. Not being excluding areas, as can be seen on [6] [7] that address human aspects and information security policy.

The critical and methodical evaluation of information security becomes necessary because business processes, technologies, and people change constantly. The challenge lies in defining information security goals, reaching them, keeping them and enhancing the controls that support them, to ensure competitiveness, profitability, compliance to legal requirements and maintaining a good image of the organization. Maturity models can help in facing this challenge [29].

Maturity strategies or models are based on the improvement of processes and the existence of fundamentals to guide and measure the implementation and improvement of processes [27][29]. There are many documents and maturity models, but, normally, it does not define a rigorous, agile, and practical maturity evaluation strategy. In many cases, the users need to build their own evaluation strategy.

Given the needs shown, this article presents strategy to measure the maturity of information security based on the ISO/IEC 27001 and 27002 standard controls according to the needs of the business environment, as well as to draw a path to demonstrate the situation of the company as subside so it is possible to prioritize the actions and resources related to the area. This adaptable strategy also uses Cobit (Control Objectives for Information and related Technology) and ISO/IEC 27005. The proposed strategy is generic and applicable to all kinds of organizations. It tries to improve the company's agility in the security field following the line of thought of Agile Governance [12][14].

The choice of COBIT, as well as the ISO/IEC 27001, 27002 and 27005 standards are due to their consolidation in the field of study.

It is believed that this analysis will contribute to the academy as, and mainly, to the industry. Since it presents, through academic research, a model for maturity and prioritization of information security that can be used by companies of different areas of activity and sizes without the need for higher investments.

This article is organized in five more sections. In the next section, the methodological design is presented. Section 3 presents the theoretical reference, ending with related works. In the fourth section, the analysis and proposed model are shown. Concluding, final considerations and opportunities for future works are presented in Section 5.

## 2. Method

According to Wohlin and Aurum [8], this research classifies itself according to what is shown in Figure 1.

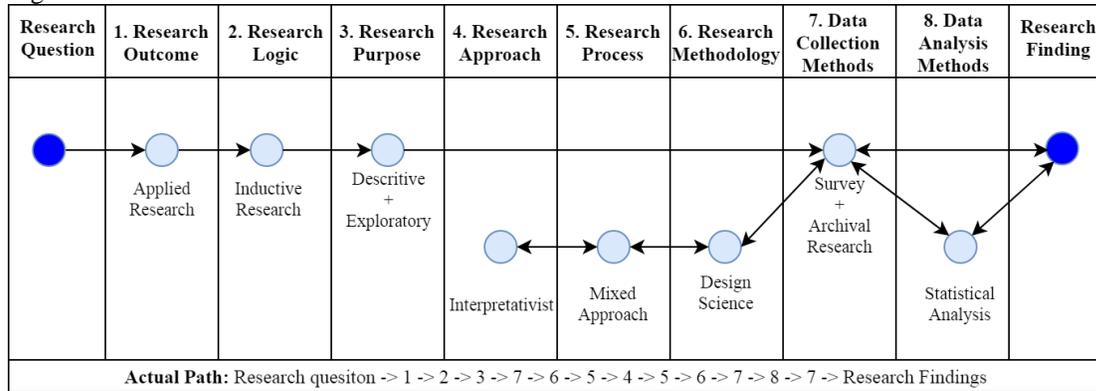

**Figure 1.** Methodological design and decision-making structure

For data collection, a survey was used, with two blocks of questions. The first one includes data of the company identification, core activity, number of employees, region of coverage, company location e other company related data. Such information was used to identify, categorize and group the companies. The second block consists in taking the 114 security controls mentioned in the ISO/IEC 27002 standard, the same ones mentioned in the Annex A of the ISO/IEC 27001 standard, and question them regarding their importance to the organization, using a Likert scale from 1 to 5 (being 1 no importance, 3 neutral and 5 very important), to select which are the most relevant. For example, below is the first question:

*Regarding the ISO/IEC 27001 and 27002 set of controls, set out below, check, from 1 to 5, how important it is in your company environment:*

*Where: 1 means no importance, 3 neutral and 5 very important*

*1. (5.1.1) It is appropriate that a set of policies for information security should be defined, approved by management, published and communicated to employees and relevant external parties.*

<div align="center">1 [ ]    2 [ ]    3 [ ]    4 [ ]    5 [ ]</div>

This method was used by other papers, for example, [9] [10].

Based on the analysis of the existing literature and obtained data, a strategy for the application of information security and measurement of maturity was proposed. For this, the average of the answers given for each control was calculated. The model was divided into four stages (using the statistic function for quartiles). Subsequently, it will be displayed if any control selected has another one of the ISO/IEC 27001 and 27002 standards as a prerequisite. If it does and it has not been selected at the same or previous stage, it will be inserted at the current stage. With the strategy set up and the controls properly inserted at each stage, the evaluation phase was initiated, and the model applied to a real case.

## 3. Theoretical Background

### 3.1 Agile Governance

Before addressing agile framework, it is important to reflect that the agile approach can be inserted in specific actions, among them the development of agile software, but it can also be applied in broader organizational context [33]. In this latter option is introduced the agile governance approach. The agile governance suggests the application of agility into the management system and organizational governance, making it more competitive [12]. Being agility, in this context, an organization's ability to react to changes in its environment more rapidly than the pace of such changes. [12].

Agility at a business level demands flexibility, responsiveness, and adaptability, which must be applied in combination with management capacity, strategic alignment, and engagement between the areas included, especially in competitive environments [12]. Agility may be understood as the capacity that organizations have, in a dynamic way, to feel the need to change from both internal and external sources and make such changes routinely without any performance loss [13]. In this regard, agile governance can be seen as the capacity of an organization to feel, adapt and respond to changes in its environment, in a coordinated and sustainable way, faster than the pace of these changes [14].

More agile means in the corporate environment, as well as with this broader vision given to agile governance, is a necessity, given that traditional Governance of Information and Communication Technology (ICT) is often a very rigid set of rules and processes, preventing the ICT from evolving and changing alongside with the company's needs and strategies [13].

Agile governance is not a replacement for the existing conventional models, frameworks, and methods (for example, ITIL and COBIT) [12]. The proposal is a new vision for governance. It is possible to compare agile governance meta-values with traditional governance, as follows [14]:

- Behavior and practice over process and procedures.
- To achieve sustainability and competitiveness over to be audited and to be compliant.
- Transparency and people's engagement to the business over monitoring and controlling.
- To sense, adapt and respond over to follow a plan.

Such bases can guide the area of corporate information security, bringing a new vision, and complementary, to the models currently proposed.

## 3.2. Information Security Governance

Information security governance (ISG) can be understood as a set of actions and practices for the alignment of information security activities with the corporate strategy [15]. ISG is a part of ICT governance, and they may overlap [16].

ISG should aim to: align the business objectives with the Information Security strategy; Ensure that the information risks are elucidated and forwarded to those responsible; as well as add value to the business, to the direction and to the stakeholders. Having as principles: establish Information Security throughout the organization; adopt a risk-based approach, recommending the joint use of ISO/IEC 27005; establish and align the investments; ensure the conformity with internal and external requirements; promote a positive security environment, including a special treatment to people; and critically analyze the performance and result of information security actions in relation with business results [17].

The creation of an effective system for assessing and managing information risks is essential [34]. Correctly applying ISG and risk analysis of information security, in addition to achieving the objectives mentioned above, tends to lead the organization to attendance and conformity with external requirements, for example, legal [16].

## 3.3. Information Security Maturity

For the correct alignment of the ICT field with the business, it is essential to set up metrics and models to define the current stage, as well as the steps to achieve a more advanced level [15] [21]. The maturity model is suitable for this purpose. A maturity model in information security is essential to obtain effective and efficient area management and governance, working together.

A maturity model is a set of characteristics, attributes, indicators or patterns that represent the capacity and progression in a determined subject. The content of the model typically exemplifies the best practices and may incorporate patterns or other subject codes of practice [22].

A model of maturity should aim to assist the companies in evaluating the information security. For the evaluation, the maturity model will indicate the current stage, as well as the path to reach more advanced levels [21]. Assisting in continuous improvement, through processes, so that the best practices can be implemented [23]. The use of a maturity model allows the identification of gaps that represent a risk and how to show them to the management team. Based on this analysis, action plans can be evaluated and developed for the improvement of processes and controls considered deficient up to the desired level of development.

Karokola, Kowalski e Yngström [24] emphasize the importance of implementing a maturity model and state that various approaches to management or maturity that support information security are already available in the market. However, it is worth mentioning that some gaps have not yet been solved and improvements can be inserted, such as the proposal to be presented in this paper.

Among the main models of maturity for information security or used in this area is the COBIT and the O-ISM3. The COBIT can be seen as a framework for the alignment between ICT and the business, that covers the best practices to be used in several areas (ICT services, Information Security, Projects, Suppliers and Software). Its controls help optimize the investment in ICT, ensure the service delivery, and provide a measure to evaluate and allow comparison. Existing its specific documents for the areas of information security [25] and risks [26]. The O-ISM3 is an information security management maturity standard published by The Open Group, the leader in the development of open and vendor-independent IT standards and certifications. The O-ISM3 standard defines security processes to manage a company's ISMS [27]. Although COBIT has a maturity model, it does not define a rigorous and practical maturity evaluation model, and the O-ISM3 does not measure risk or security directly [29]. Another difference between this work and COBIT and O-ISM3 is that the COBIT and O-ISM3 see all the controls with the same criticality.

## 3.4. ISO/IEC 27001 and 27002

For the effective management of information risks and security of information, special techniques have been developed, for example, the methodology of international standards [34], for example, the ISO/IEC family 27000. The ISO/IEC 27001:2013 standard points out the basic requirements for the implementation of an Information Security Management System (ISMS), as well as all of its controls and management [18]. It is the main standard that an organization should use as a basis to obtain the business certification in information security management. Therefore, it is known as the only international standard that can be auditable and that defines the requirements for an ISMS. It consists, in its current version, of 14 sections of information security controls, 35 control objectives and 114 controls that can be implemented [18].

The ISO/IEC 27002 standard [19] can be seen as a complement to the ISO/IEC 27001. It contemplates the same division of the ISO/IEC 27001, 14 sections of information security controls, 35 control objectives and 114 controls that can be implemented [18], being a detailing of the implementation. An example of this structure: Section 12 - Operations security. 12.3 Backup - Objective: To protect against loss of data. 12.3.1 Information backup - Control: Backup copies of information, software and system images should be taken and tested regularly in accordance with an agreed backup policy.

It can be understood as a code of practices with a complete set of controls that aid the application of the Information Security Management System. It is recommended that the standard should be used in conjunction with the ISO/IEC 27001, but it may also be consulted independently for the adoption of good practices. The identification of which controls should be implemented requires careful planning and attention to detail. It is highlighted as one of the main benefits of this model the prevention against the financial loss that the organization may have in the event of information security incidents [20].

These standards (ISO/IEC 27001 and 27002) are internationally recognized as one of the main frameworks in the field of information security, listed as one of the main reference documents for the elaboration of an ISMS and an information security policy.

## 3.5. Related Works

When a model of maturity is used based on a model of best practices, these models achieve a maturity measurement with considerable and grounded results, for example, as occurs with a set of works, for instance, with the COBIT Management Guidelines, the Process Maturity Framework (PMF) [23] and the present research.

In addition to the COBIT and the O-ISM3, already presented, it is highlighted the Cybersecurity Framework from the National Institute of Standards and Technology (NIST) [28] that, even without a maturity model, is a relevant document for the area. The framework consists of standards, guidelines, and best practices to manage cybersecurity-related risks in an attempt to promote the protection and resilience of the critical infrastructure and other important sectors to the US Government.

Among the existing academic works, in this same field of study, it is possible to cite the cases of the cyclic evaluation model of information security maturity proposed by Rigon *et al.* [29], which is derived, among other concepts, from the ISO/IEC 27002, ISO/IEC 27005 and the COBIT; as well as the studies of Gomes *et al.* [23] and Park *et al.* [30] that bring results of ITIL-based maturity models.

Based on the ISO/IEC 27001 it is given the information security maturity model proposed by [21], a solution focused on software development companies. Based on the ISO/IEC 27002 and the Systems Security Engineering Capability Maturity Model (SSE-CMM) it is given the information security maturity model proposed by [20]. The GAIA-MLIS [31] is an information security maturity model based on the COBIT 5 and ISO/IEC 27001 and 27002 standards. It has five maturity levels ranging from 0 to 4, evaluating 5 distinct areas (hardware, software, installations, personnel, and information).

The studies cited correlate with the present research because they try, to a certain extent, to measure information security based on a framework already consolidated in the market. Other difference lies in the fact that the studies cited present proposals that are static and with the same value to all the controls. This type of analysis is not appropriated to the business in a constant evolution of the environment. Using them as base, the current paper can be seen as an evolution of the mentioned models and proposes not only a measurement model, but also a guide to the implementation of information security, according to the importance pointed out by the companies, shaped as a modular strategy that allows the use of the various existing frameworks.

## 4. Proposed Strategy

The adaptable strategy consists of six steps (Figure 2). Initially, in raising the company's view on the criticality of each one of the 114 ISO/IEC 27001 (Annex I) and 27002 standard controls, scoring in a scale from 1 (less important) to 5 (very important). After the response, in a second step, the controls will be sorted into quartiles. For this, the average of the scores of each control will be calculated and ordered.

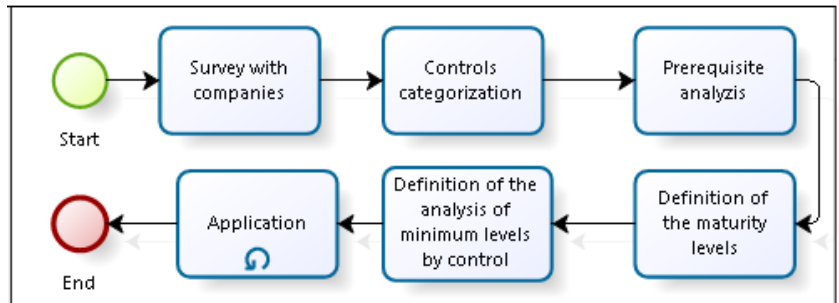

**Figure 2**. Macro steps of the strategy

The first quartile represents 25% of the controls considered most important, while the last quartile will indicate the 25% of the controls with the lower level of importance. Each quartile is categorized as "stage". In an ideal scenario, the controls would be distributed as shown in Table 1.

**Table 1.** Division of the controls by quartiles

| Average of the Control | 1st Quartile (highest averages) | 2nd Quartile | 3rd Quartile | 4th Quartile (lowest averages) |
|---|---|---|---|---|
| Stage | Essential | Intermediate | Advanced | Full |
| Controls | 29 | 28 | 29 | 28 |

However, the division of the quartiles may not be as accurate as proposed (Table 1), given that after the first division of the quartiles it should be analyzed if there is any dependency or prerequisite between controls. If there is, the prerequisite control should be in the same quartile or in a previous one. If the control is in a latter quartile, the prerequisite control will be placed in the same quartile as its dependent (3rd step). For example, a situation where, initially, the ISO/IEC 27001 control "A.5.1.1 - A set of policies for information security shall be defined, approved by management, published and communicated to employees and relevant external parties" is in the 3rd quartile, while the ISO/IEC 27001 control "A.5.1.2 - The policies for information security shall be reviewed at planned intervals or if significant changes occur to ensure their continuing suitability, adequacy and effectiveness. " is in the 2nd quartile. It is noticed that the control A5.1.1 (definition of the information security policy) is a prerequisite to the control A.5.1.2 (analysis of information security policy), in which case the control A.5.1.1 will be inserted in the 2nd quartile, the same as control A.5.1.2.

A quartile may also have its number of controls increased if there is a tie in the scores of the last controls, all of which are incorporated into the quartile. For example, if the controls of positions 27, 28, 29 and 30 have the same average, all will be incorporated into the 1st quartile, having, in this case, the 1st quartile 31 controls instead of the initial 29 (Table 1).

The fourth step consists in defining the maturity levels. This proposal uses the maturity levels and definitions of COBIT [32], which are:

- **Level 0 (Non-existent):** Complete lack of any recognizable processes. The enterprise has not even recognized that there is an issue to be addressed.

- **Level 1 (Initial/Ad Hoc):** There is evidence that the enterprise has recognized that the issues exist and need to be addressed. There are, however, no standardized processes; instead, there are ad hoc approaches that tend to be applied on an individual or case-by-case basis. The overall approach to management is disorganized.

- **Level 2 (Repeatable but Intuitive):** Processes have developed to the stage where similar procedures are followed by different people undertaking the same task. There is no formal training or communication of standard procedures, and responsibility is left to the individual. There is a high degree of reliance on the knowledge of individuals and, therefore, errors are likely.

- **Level 3 (Defined Process):** Procedures have been standardized and documented and communicated through training. It is mandated that these processes should be followed; however, it is unlikely that deviations will be detected. The procedures themselves are not sophisticated but are the formalization of existing practices.

- **Level 4 (Managed and Measurable):** Management monitors and measures compliance with procedures and takes action where processes appear not to be working effectively. Processes are under constant improvement and provide good practice. Automation and tools are used in a limited or fragmented way.

- **Level 5 (Optimized):** Processes have been refined to a level of good practice, based on the results of continuous improvement and maturity modelling with other enterprises. IT is used in an integrated way to automate the workflow, providing tools to improve quality and effectiveness, making the enterprise quick to adapt.

The fifth step is to define the minimum maturity level for each control. One way is to list a pattern, for example, all of the controls must reach at least level 3 (defined process). Another option, more recommended, is to perform a risk analysis in the company, as suggested by the ISO/IEC 27005 standard, and categorize the minimum level of each control according to its probability and impact.

For this strategy, the probability and impact are categorized as low, medium or high. Each will receive, respectively, the weights 1, 2 or 3. A matrix is created and the minimum maturity value to be reached will be the sum of the probability and impact scores, according to Figure 3.

| | High (3) | 4 - Managed | 5 - Optimised | 6 - Optimised |
|---|---|---|---|---|
| **Probability** | Medium (2) | 3 – Defined | 4 - Managed | 5 - Optimised |
| | Low (1) | 2 – Repeatable | 3 – Defined | 4 - Managed |
| | | Low (1) | Medium (2) | High (3) |
| | | **Impact** | | |

**Figure 3.** The minimum level of maturity according to impact and probability

An exception is when it is reached a high probability and impact, receiving a score of 6 (3+3). Aware that the proposed model addresses maturity levels up to maturity level 5 (optimized), the controls categorized with a score of 6 must reach level 5 (optimized) and, due to their criticality, must be treated as a priority by the company.

This way, the controls which, in case of its absence, creates risks with greater probability and impact should be treated differently, with a greater maturity level. The controls that are not applicable should be properly justified in the report to be presented at the end of the evaluation and will have the minimum level 0 (non-existent), not being counted in the strategy.

After the initial definitions, the application of the strategy is set as the sixth step. The strategy implementation phase consists of analyzing and applying each control ordered in the stages up until the desired stage, according to Figure 4.

It is emphasized that achieving more advanced levels and stages of maturity generate costs and demand time, and it may not be interesting for all companies to reach the complete stage and the maturity level 5 (Optimized) in all of its controls.

It is also important to mention the final application phase, which consists of generating the Evaluation and Stakeholders Communication Report of the analyzes made, levels to be achieved, strengths and weaknesses of the company. The whole strategy must be repeated periodically, creating a PDCA cycle (Plan, Do, Check, Act).

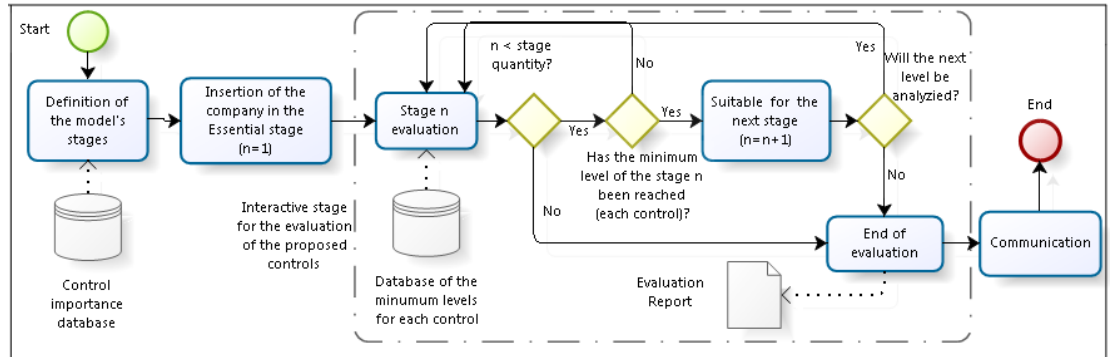

**Figure 4.** Strategy implementation phase

## 4.1 Security Maturity Model

To be able to compare the companies and measure their maturity, a standard model is necessary to compare the same issues. Information security maturity models, such as those cited in the Related Works section, use, for example, the ISO/IEC 27001 or 27002 standards controls by measuring them and classifying the maturity, usually, with the average of the measured values. All controls have the same weight.

As a differential of this adaptable strategy, the proposed maturity model works with stages and levels of maturity. In the stages are classified the controls by quartiles, according to the importance given to them by the companies. The company will only be evaluated in the second quartile by meeting the minimum level of controls of the previous stage. In this new configuration, the most important controls will be inserted as a priority. The proposed maturity model has its stages and levels shown in Figure 5.

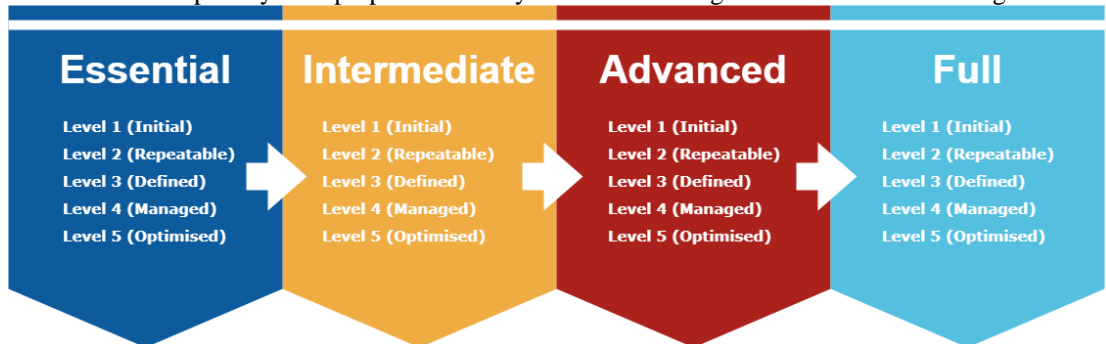

**Figure 5.** Stages and levels of the maturity model

In the proposed model, to go from one stage (Essential, Intermediate, Advanced or Full) to the next one is only possible after reaching the minimum maturity level of 3 for all controls, this being the parameter of the minimum levels Database for each Control (Figure 4). That is, if the company is categorized as Advanced Level 2, it means that the average of the controls in the Advanced group obtained maturity level 2 (Repeatable) and that all applicable controls of the Essential and Intermediate stages were measured at least as level 3 (Defined). Being a differential of the proposed model.

In order to verify which ISO/IEC 27001 and 27002 controls are the most important, a survey with Brazilian companies were used. Answers from 157 different companies were obtained, with 23 being multinational, as described in Figure 6. The Survey was answered by ICT professionals or with training in the field. It is noteworthy that 93 of them (60.78%) worked exclusively or primarily with the information security area in their respective companies.

Based on the average of the responses given by the companies reached, the controls were divided between the stages, reaching the configuration shown in Table 2, which represents the control importance database (Figure 4). The numbering of the controls is in accordance with what is shown in Annex A of the ISO/IEC 27001 standard.

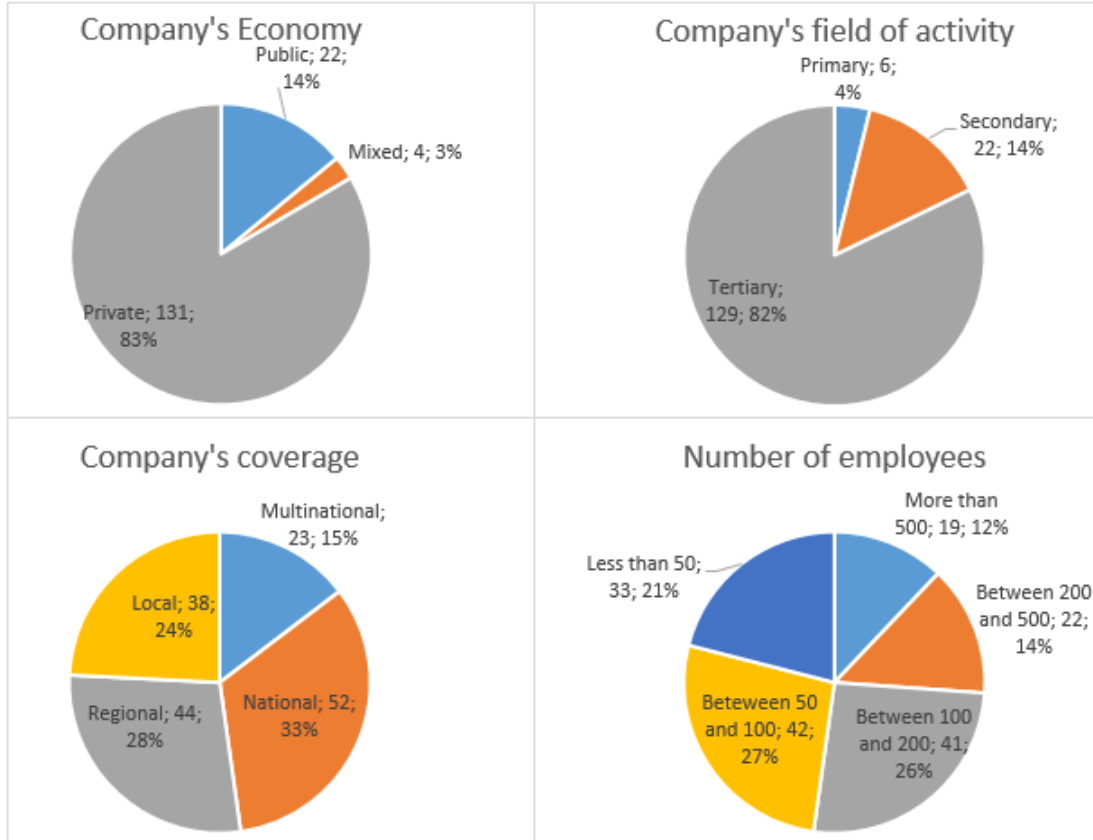

**Figure 6.** Description of the sample achieved

The controls of each stage are recalculated periodically, due to the possibility of new answers to the model, increasing the database and representing the real importance of each control. This fact confers a dynamism to the model, being another differential.

**Table 2.** Division of controls by quartiles

| Stage | Number of Controls | Controls |
|---|---|---|
| **Essential** | 31 | A.5.1.1, A.6.1.1, A.6.1.5, A.6.2.2, A.7.1.1, A.7.2.1, A.8.1.2, A.8.1.3, A.8.2.1, A.8.2.3, A.9.1.2, A.9.2.1, A.9.2.3, A.9.2.4, A.9.2.5, A.9.4.2, A.9.4.4, A.11.1.5, A.11.2.4, A.11.2.5, A.11.2.6, A.11.2.7, A.12.5.1, A.12.6.2, A.13.1.3, A.15.1.3, A.18.1.1, A.18.1.2, A.18.1.3, A.18.1.4 e A.18.1.5 |
| **Intermediate** | 27 | A.5.1.2, A.6.1.2, A.6.2.1, A.7.2.2, A.8.1.1, A.8.3.1, A.9.2.6, A.9.4.3, A.11.1.3, A.11.2.2, A.11.2.3, A.12.1.3, A.12.1.4, A.12.2.1, A.13.1.1, A.13.2.1, A.13.2.3, A.14.1.1, A.14.2.6, A.15.1.1, A.16.1.1, A.16.1.2, A.16.1.4, A.16.1.5, A.16.1.7, A.17.1.1 e A.18.2.2. |
| **Advanced** | 29 | A.7.2.3, A.8.1.4, A.8.2.2, A.9.1.1, A.9.3.1, A.9.4.1, A.9.4.5, A.11.1.1, A.11.1.2, A.11.2.1, A.11.2.9, A.12.1.1, A.12.1.2, A.12.3.1, A.12.4.1, A.12.6.1, A.12.7.1, A.13.1.2, A.13.2.4, A.14.1.2, A.14.1.3, A.14.2.5, A.14.2.9, A.15.1.2, A.15.2.1, A.16.1.3, A.17.2.1, A.18.2.1 e A.18.2.3. |

| **Full** | 27 | A.6.1.3, A.6.1.4, A.7.1.2, A.7.3.1, A.8.3.2, A.8.3.3, A.9.2.2, A.10.1.1, A.10.1.2, A.11.1.4, A.11.1.6, A.11.2.8, A.12.4.2, A.12.4.3, A.12.4.4, A.13.2.2, A.14.2.1, A.14.2.2, A.14.2.3, A.14.2.4, A.14.2.7, A.14.2.8, A.14.3.1, 15.2.2, A.16.1.6, A.17.1.2 e A.17.1.3. |

## 4.2 Independent Application of the Strategy

The independent application of the adaptable strategy aims to assist companies that are searching for the implementation of information security and need a strategy that suits their company and does not need, at this moment, to use the best practices pointed out in the maturity model (Section 4.1).

It consists of applying the previously mentioned strategy (Figures 4, 5 and 6). However, the Controls Importance Database (Figure 4) will not be the one presented in Table 2, but rather an analysis of the application of the same Survey (questioning the importance of each control) within the company itself with the Stakeholders.

The minimum level Database for each control (Figure 4) will be set up according to the risk assessment for each control, according to the matrix presented in Figure 3.

In this way, we have the application of the same strategy but adapted to the environment and needs of the company. This fact follows the line of thought of Agile Governance, especially its meta-values that point [14]: (i) Behavior and practice over process and procedures, (ii) To achieve sustainability and competitiveness over to be audited and to be compliant, (iii) Transparency and people's engagement to the business over monitoring and controlling, and (iv) To sense, adapt and respond over to follow a plan. Factors that are happening in the process.

In this application the company is shaping the process to its business and not the opposite to be compared with others, aiming to achieve their goals, prioritize their actions in information security, as well as to promote basic definitions for the treatment of controls, having parameters for definition of investment and functioning with an alignment of information security to the business, which can be seen as Information Security Governance, which supports ICT Governance, which, finally, incorporates Corporate Governance.

## 4.3 Real Application of the Strategy

To verify the functionality of the proposed strategy, a real application of the maturity strategy was proposed with a Brazilian Company, presented in this paper as Company A. In contact with the Company A, it opted for the independent application of the strategy (Section 4.2).

The survey to define the importance of each control was answered by seven employees, three of them from the ICT area and four managers and directors from other areas. With the answers, the average importance of each control was calculated, following the same methodology.

The division of controls by levels indicated a change in eleven controls when compared with the base exposed in Table 2. Three controls by exclusion (not applicable to the company), being: "A.14.2.1 - Secure development policy", "A. 14.2.6 - Secure Development Environment "and" A.14.2.7 - Outsourced Development" and 8 by different categorization, as detailed in Table 3.

**Table 3.** Company A x Control Database

| Control | Stage in the Database (Table 2) | Stage in the Company A Database |
|---------|--------------------------------|-------------------------------|
| A.5.1.2 | Intermediate | Essential |
| A.6.1.2 | Intermediate | Essential |
| A.6.1.5 | Essential | Intermediate |
| A.6.2.2 | Essential | Full |
| A7.1.1 | Essential | Advanced |
| A.7.2.3 | Advanced | Full |
| A.11.1.2 | Advanced | Intermediate |
| A.18.1.4 | Essential | Intermediate |

With the exposed alterations the company's strategy resulted in 29 controls in the Essential stage, 27 in the Intermediate, 28 in the Advanced and 27 in the Full.

Four months earlier, the Company A had carried out an internal risk analysis based on the ISO/IEC 27005 standard. The report of this action served as input for respondents to define the score of each control and, mainly, to define the minimum level of each control, according to the proposed matrix (Figure 3). With the two Databases formed (controls and minimum level), the adaptable strategy was followed, according to Figures 4 and 5.

The company met all the minimum criteria of the Essential controls, resulting in a maturity measured at the stage of 3.49, being able to move to the next stage. In the Intermediate controls, the analysis resulted in an average maturity of 3.30, having met the minimum level in all controls, and qualifying to the next stage. In the Advanced stage, the company reached an average maturity of 3.07. However, it did not reach the minimum level in two controls: "A.9.3.1 - Use of secret authentication information", which reached level 3 (Defined) and the Risks matrix had indicated with minimum criteria level 4 (Managed), and the control "A.14.1.3 - Protecting application services transactions", which reached level 2 (Repeatable) and the Risks matrix had also pointed to level 4 (Managed) with a minimum criterion.

By not having completed the Advanced Placement, Company A had its general maturity level defined as Intermediate Stage, Maturity Level 3.30 (Defined).

The company opted for the independent application of the adaptable strategy (Section 4.2) because it did not have an institutional need for certification of its maturity or to compare it with other companies. Adding that the independent strategy allowed the Company to design a way to direct its resources to the controls most critical to its business, a fact that was evidenced in the Stakeholder Communication Evaluation Report, since controls of the Full stage, that is, not categorized as critic to business, pointed to a maturity level 5, while some previous, more critical, stage controls were at level 2 (repeatable). Demonstrating that, possibly, efforts and resources were directed to non-priority actions at the time.

If the company opted for the Maturity Model strategy (Section 4.1), analyzed according to the description of the controls in Table 2, it would be categorized in the Essential Stage, with Maturity Level 3.42 (Defined). Even though the company obtained maturity levels 5 (Optimized) in some controls, including in the last stage (Full), the company had two controls with maturity level 2 (Repeatable) in the Intermediate stage, not reaching the minimum level 3 and, therefore, not completing this second stage.

By making a general average of the Controls, as it is done in most of the works, including in the related works, Company A would be inserted with maturity level 3.22, Defined Level. Although it is the same level indicated by the two strategies proposed in this work, it is believed that the present proposal is richer, since it points out differences for the same level of maturity. Making it clear that an Essential stage company, 3.3 maturity level is at a level well below a Full stage company, maturity level 3.3. Even if both can have the same level of maturity if calculated by the traditional molds (simple average of the levels of each control), being in a more advanced stage, points to the attendance of the main controls to the business.

## 5. Final Considerations

In the current environment, it is very clear the necessity to seek improvements in information security, as well as means to ensure the best use of time and resources. In several cases, such actions can be a differential for the continuity of the business.

The work presented an adaptable strategy for information security that can be applied as a model of maturity, as well as being possible its independent application focused on am alignment between the business and information security controls, being successfully tested in a company. This strategy is applicable to all kinds of organizations

The adaptable strategy allows a more improved analysis of the maturity level, giving greater visibility through the stages, if the main controls were measured and reached their minimum level, not only analyzing a general level of maturity of the company, which can be achieved by meeting requirements that are easier or not so critical to the business. As well as analyzing the final report, adjusting and applying the strategy periodically, creating a PDCA cycle.

The proposed strategy basically consists of a set of modules: The 4 stages proposed (Essential, Intermediate, Advanced and Full), maturity levels (based on COBIT), the aspects and controls to analyze (using the ISO/IEC 27001 and 27002 standard controls) and the definition of the minimum level of each control, pre-defined or based on the inherent risk of each one (ISO/IEC 27005). It is believed that this structure is a differential of the proposed strategy, since it uses consolidated frameworks, but also allows, in the independent application (Section 4.2), the exchange of some module by another at the discretion of the company. For example, a set of different maturity levels already used in the company or a risk analysis based on other parameters. This possibility allows greater adjustment of the adaptable strategy to the business, as well as reducing time and resources in your application by using some already known and existing module.

Being able to evaluate as an independent application, using modules, allowing adjustments to the business, points the convergence to the thoughts of agile governance.

The use of this strategy can be inadequate for beginner users, can inserting not appropriated controls in the stages (not critical in the first stages and critical in the lasts). This hypothesis needs to be tested. However, for not beginner users, the strategy provides several adjustments and expansions of the scope of evaluation changing according to the business goals.

As a continuation of the work in the future, it is visualized a survey with companies to create the base of a minimum level of controls through the risk matrix (impact x probability) to be used in the Maturity Model as a minimum level for each control instead of the fixed level 3. Keeping the base of controls and minimum level by inserting new companies, generating a basis of "best practices" and updating the model in a dynamic way, as well as having sufficient quantity to analyze or generate different bases for specific niches, according to the geographical region (country, continent), organization size, and market type (Industry, trade).